\documentclass[12pt,a4paper]{article}
\usepackage[utf8]{inputenc}
\usepackage[english]{babel}
\usepackage{amsmath}
\usepackage{amsfonts}
\usepackage{amssymb}
\usepackage{graphicx}
\usepackage{graphics}
\usepackage{float}
\usepackage[algosection,ruled,lined,linesnumbered,longend]{algorithm2e}
\usepackage{hyperref}
\usepackage{cleveref}

\usepackage{authblk}

\title{On the Deformed Algebras Applications to Quantum Physics}
\author{Julio Cesar Jaramillo Quiceno\thanks{jcjaramilloq@unal.edu.co}, Plamen Neytchev Nechev\thanks{pnechev@abv.bg}}
\affil{
 \textsuperscript{1}Departamento de Física, Universidad Nacional de Colombia, Edificio Yu Takeuchi, Bogotá
  \textsuperscript{2}Independent Researcher, Strasbourg, France.
}
\date{}
\begin{document}
\maketitle
\begin{abstract}
Some possible applications of deformed algebras to Quantum Physics are considered based on a rigorous approach. Jackson integrals are expressed in the context of the equipped separable Hilbert space. Jackson integrals are expressed in the context of the equipped separable Hilbert spaces using point measures where possible. Along the way, certain errors and/or inaccuracies made by different authors of the cited references have been corrected. A brief analysis at the end of the article indicates that there are still problems in applying deformed algebras to Quantum Mechanics and Field Theory.
\end{abstract}
\subsubsection*{Keywords:} Deformed Algebras, Jackson Calculus, Equipped Hilbert Space, Lattices 
\section{Introduction}
It is common knowledge that the Quantum Physics has had many successes, and its concepts and usages tower over many scientific applications and practically all the modern technology. However, there are still a big number of inconsistencies, when considering certain branches of its development such as for example, Field Theories. In this article it is attempted to evince the mathematical structures called deformed algebras, therewith improving certain aspects. Historically, the Matrix Mechanics, was developed by Heisenberg (1925) \cite{Heis-25, Zetilli-09} to describe atomic structure starting from observed spectral lines. With the development of mathematics, it became evident that the operators (or matrices) introduced by Heisenberg can be considered as a basis for constructing certain free algebras. This allowed us to go in the opposite direction, defining different algebraic structures (called algebras) and trying to apply them to solve some problems in Quantum Physics. In \cite{schmudgen-09}, the $q$-Heisenberg algebra is defined for a real positive number$q\neq 1$. One year later, Wess in \cite{Wess-00} defined the commutation relations for the q-deformed Heisenberg algebra, in a different way. Silvestrov and Hellström introduced another version of the q-deformed Heisenberg algebra in \cite{Silvestrov-Hellstrom-00}, motivated by the well-known Heisenberg canonical relation $\hat{x}\hat{p}-\hat{p}\hat{x}=I$ where $I$ denotes the identity operator. This relation occupies a key place in the operator formulation of quantum mechanics. The same authors also mention the physical motivation behind this deformation, which arises from the creation and annihilation operators for systems with one degree of freedom and subject to Bose statistics. Other deformations are proposed by Wess and Schwenk \cite{wess-schwenk-1999} in the context of the quantum plane \cite{jordan-wells-1996, Chen-Lu-2021} and, at present, there exists the so-called quantum calculus. The $q$-derivative and $q$-integral has been proposed in references \cite{lavagno-gervino-09, Jackson-1909}. The next chapter is dedicated to a very brief description of the $q$-deformed algebras. In chapter three, the Jackson derivative is defined, considering its use in Quantum Physics. It follows the short presentation of the equipped Hilbert spaces conception and its relationship with the new "deformation" notions. In the fifth chapter are considered the corresponding integral calculus where, the similarities and differences with the respective formulae based on known traditional calculus definitions can be appreciated. Since all the physics laws are closely associated to the existence and uniqueness theorems of differential equations, within the penultimate chapter an attempt is made to extend these theorems to the already introduced strain calculus. The article concludes with some discussions and conclusions regarding the applicability of the deformed algebras in Quantum Physics.

\section{Deformed Algebras}

According to the Heisenberg quantum mechanical picture, the three position and impulse operators beside the identity one, are the base of so-called Heisenberg (traditional) algebra. The \( q \)-deformed Heisenberg algebra or \( q \)-Heisenberg algebra has been defined and in detail studied by various authors such as \cite{Wess-00, schmudgen-09, wess-schwenk-1999, Silvestrov-Hellstrom-00, lavagno-gervino-09}. The \( q \)-deformed Heisenberg algebra is determined by commutations between the coordinates of any by the impulses of the different projections and the following commutation relations for the different type of magnitudes corresponding to the same projections (for \( q \in \mathbb{R} \)):

\[
[\hat{x}, \hat{p}_x] = [\hat{y}, \hat{p}_y] = [\hat{z}, \hat{p}_z] = i f(q).
\]

Now, if \( q \in \{-1, +1\} \) and \( f(q) = -i \hbar^{-1} \), then this definition can be obviously extended to the definition of the generalized \( q \)-Heisenberg algebra as shown in \cite{Silvestrov-Hellstrom-00}; if \( q \in \mathbb{C} - \{0\} \), and \( f(q) = i \hbar D_{p_k}(q) \), being \( D_{p_k}(q) \) any function that depends on \( q \), results the \( q-\hbar \) Heisenberg algebra proposed in \cite{Volovich-Arefeva-91}. In the last case, two deformation parameters are used: \( q \) and \( \hbar \) i.e. the Planck "constant" is a variable. 

With respect to the harmonic oscillator, let \( \hat{a}, \hat{a}^\dagger \) be annihilation and creation operators and \( N \) the quantum number operator. The \( q \)-Heisenberg algebra is defined by the generators \( \hat{a}, \hat{a}^\dagger \) and \( \hat{N} \) subject to the following relations \cite{lavagno-gervino-09}:

\[
\hat{a}\hat{a}^\dagger - q\hat{a}^\dagger \hat{a} = q^{-\hat{N}}, \quad [\hat{a}, \hat{a}^\dagger] = [\hat{a}^\dagger, \hat{a}^\dagger] = 0, \quad [\hat{N}, \hat{a}^\dagger] = \hat{a}^\dagger, \quad [\hat{N}, \hat{a}] = -\hat{a}.
\]

For greater comfort let us introduce the following two symbols [...]:

\[
[n]_q = \frac{q^n - q^{-n}}{q - q^{-1}},
\]

and the "factorial"

\[
[n]_q! = [n]_q [n-1]_q \cdots [1]_q,
\]

being \( q \) a non-zero real number. We can put \( [0]_q = 0 \) and also calculate that obviously \( [1]_q = 1 \). But unfortunately, \( [n]_q \neq n \) for any other natural number. The above notation makes easier to express different types of operators through formulae that supplant the common concepts of differential and integral calculus. In what follows, the considered operators are assumed to set on the elements of appropriate vector spaces.

\section{Jackson Derivative}

The definition of the Jackson derivative is the following:

\begin{equation}\label{Jackson-Der}
D^{(q)}_x f(x) = \frac{f(qx) - f(q^{-1}x)}{(q - q^{-1})x} ,
\end{equation}

where the deformation parameter \( q \in (0, 1) \). The above formula is symmetric with respect to the transformation \( q \mapsto q^{-1} \). If the function \( f(x) \) satisfies the usual differentiability conditions, substituting \( q x \) by \( x + \Delta x \), we can make sure that the new derivative consists of the usual limit when \( q \to 1 \). The eq. \eqref{Jackson-Der} defines a "finite" increasing derivative and would make sense if the deformation \( q \) is remarkably close to 1. The last considerations, together with some relations in the next chapters, would justify the introduction of the said expression as Jackson derivative. Immediately we can demonstrate several properties of \( D^{(q)}_x \) \cite{Kac-2002}:

\begin{align}
\label{D1}D^{(q)}_x x^n & = \  [n]_q x^{n-1} ,\\
\label{D2}D^{(q)}_x (f(x)g(x)) & =\  D^{(q)}_x [f(x)] g(q^{-1}x) + f(qx) D^{(q)}_x [g(x)] \\
\notag & = \  D^{(q)}_x [f(x)] g(qx) + f(q^{-1}x) D^{(q)}_x [g(x)] ,\\
\label{D3}D^{(q)}_x [\alpha f(x) + \beta g(x)] &  = \  \alpha D^{(q)}_x [f(x)] + \beta D^{(q)}_x [g(x)] ,
\end{align}

(\(\alpha\) and \(\beta\) are assumed to be constant)

\begin{align}
\label{rel1}[\hat{x},\hat{y}] & = \  [\hat{y}, \hat{z}] = [\hat{z}, \hat{x}] = 0 ,\\
\label{rel2}[\hat{p}_x, \hat{p}_y] & =\  [\hat{p}_y, \hat{p}_z] = [\hat{p}_z, \hat{p}_x] = 0,
\end{align}

and

\begin{equation}
[\hat{x}, \hat{p}_x] = [\hat{y}, \hat{p}_y] = [\hat{z}, \hat{p}_z] = i \hbar  .
\end{equation}

Here, once fixing \( q \), the position and impulse operators are defined as follows: \( \hat{x} = x \), \( \hat{y} = y \) and \( \hat{z} = z \); respectively \( \hat{p}_x = -i \hbar D^{(q)}_x \), \( \hat{p}_y = -i \hbar D^{(q)}_y \) and \( \hat{p}_z = -i \hbar D^{(q)}_z \). The symbol \([... , ...]\) (different from \([... , ...]_q\)) is the current commutator defined in the ordinary Quantum Physics.

\section{Equipped spaces}

Let \( H_z \) be the Hilbert space consisting of the equivalence classes of the Lebesgue-integrable functions \( (\psi \in \mathbb{C}) \) on the interval \((-\infty, \infty)\), \( (H_z \approx L_z^2) \) because all the separable Hilbert spaces are isomorphic) i.e. the condition of belonging is the request for integrability
\[
\int_{\mathbb{R}} |\psi(x)|^2 \mu(dx) < \infty ,
\]
and \( (\mathcal{S}^* \subset H_z^*, \mathcal{S}) \) be its respective equipped space. Since the \( \mathcal{S} \) functions are continuous, in the corresponding equivalence classes created by the functions of \( H_z \), only one \( \mathcal{S} \) representative enters. Note, that the Dirac delta function (actually, it is the functional \( \delta_{x_0} \)) can be realized with a point measure \( \mu^* \) adopted at \( x_0 \):
\[
\delta_{x_0}(f) = \int_{-\infty}^{\infty} f(x) \mu^*(dx) = f(x_0),
\]
being \( f \in \mathcal{S} \). This extends over any finite sum.

Afterward, the space \( H_z \) is separable and one can construct a Fock basis starting from a vacuum function \( \psi_0(x) \) defined by the destruction or annihilation operator \( D^{(q)}_x \):
\begin{equation}\label{psi-0}
D^{(q)}_x \psi_0(x) = 0.
\end{equation}
The role of the creation operator is fulfilled by \( \hat{x}\). In fact, let us introduce the function \( \psi_n(x) = \frac{1}{\sqrt{[n]_q!}} \hat{x}^n \psi_0(x) \) belonging to a Fock base (although the normalization loses its initial traditional physical meaning). Then,
\[
D^{(q)}_x \psi_n(x) = \sqrt{[n]_q} \psi_{n-1}(x),
\]
and
\[
\hat{x} \psi_n(x) = \sqrt{[n+1]_q} \psi_{n+1}(x),
\]
being \( n = 0 \) or \( n \in \mathbb{N} \). Thus, for the operator regarding the quanta number ("how many quanta" answering) we have
\begin{equation}\label{N-n}
\hat{N} \psi_n(x) = \hat{x} D^{(q)}_x \psi_n(x) = [n]_q \psi_n(x).
\end{equation}
Here, we differ from the publications \cite{Vladimirov-1971}, where the \( [n]_q \) numerical value is consciously or unconsciously overlooked.

\section{Deformed integrals}

In the reference \cite{Kac-2002}, a deformed or Jackson integral is defined in the following manner:

\[
\int_0^b f(x) d_q x = b(q^{-1} - q) \sum_{n=0}^{\infty} q^{2n+1} \delta_{2n+1}(f) = b(q^{-1} - q) \sum_{n=0}^{\infty} q^{2n+1} \delta_{2n+1}(f),
\]

(for the second equality it is necessary to suppose \( f \in \mathcal{S} \), we disagree with \cite{Vladimirov-1971}, page 115), i.e., as a point measure too, who is a non-negative number. For the right-hand side of the above equation to make sense without limiting the functions too much, we will assume that \( |f(x)| \leq 1 \) (or any \( \leq M > 0 \)). Additionally, if \( f \in \mathcal{S} \) we also can define an infinite deformed integral:

\[
\int_0^{\infty} f(x) d_q x = (q^{-1} - q) \sum_{n=0}^{\infty} q^{2n+1} f(q^{2n+1}) = (q^{-1} - q) \sum_{n=0}^{\infty} q^{2n+1} \delta_{2n+1}(f),
\]

using the behaviour of \( f \) when \( b q^{-2n+1} \) and in divergent word to... It is advisable to continue the analogy with current integral calculus; let us put

\begin{equation}\label{integral}
\int_{a}^{b} f(x) d_q x = \int_{0}^{b} f(x) d_q x - \int_{0}^{a} f(x) d_q x.
\end{equation}

Unfortunately, from the following contra-example (a and b are close enough, \( q = 1/2 \)):

\begin{figure}[H]
    \centering
    \includegraphics[width=0.8\textwidth]{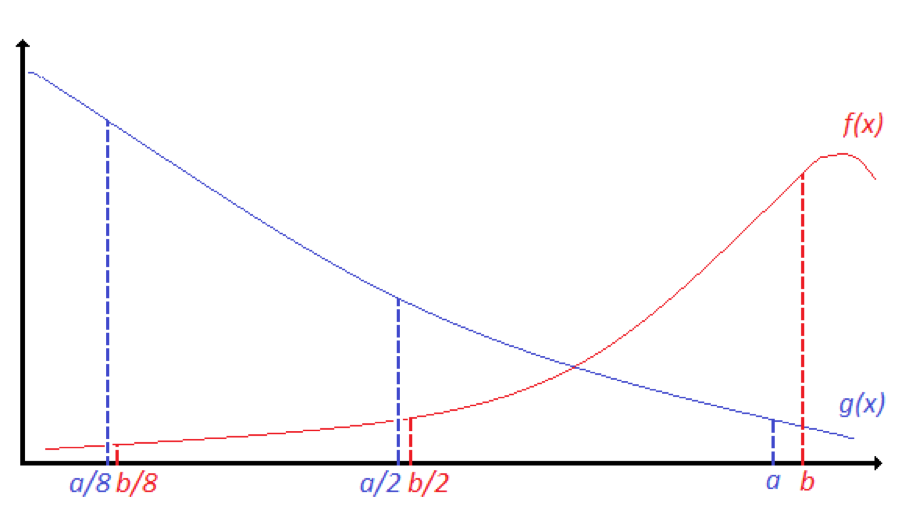}
    \caption{Example of deformed integral comparison}
    \label{fig:deformed_integral}
\end{figure}

One can see that \( f(x) > g(x) \) on the interval \([a,b]\) does not implicate mandatory

\[
\int_{a}^{b} f(x) d_q x > \int_{a}^{b} g(x) d_q x.
\]

The latter points that not all introduced here integrals are expressible by any point measures and therefore, the integrals outlined shown in the Stenberg \cite{Stenberg-1962}  which allows stable its uniqueness is inapplicable.

If the definition of the improper integral at \(\infty\), where \( d_q x \) can be considered as a measure, is extended toward \(-\infty\) by the natural assumption that \( d_q(-x) = - d_q x \), we obtain:

\[
\int_{-\infty}^{\infty} f(x) d_q x = (q^{-1} - q) \left( \sum_{n=-\infty}^{\infty} q^{2n+1} f(-q^{2n+1}) + \sum_{n=-\infty}^{\infty} q^{2n+1} f(q^{2n+1}) \right),
\]

believing that at this point the first series convergence is obvious (\( f \in \mathcal{S} \)). In reality, all the allowed functions \( f \) have contributions only on the following countable set of points:

\[
..., -q^{-2n+1}, ... -q^{-3}, -q^{-1}, -q^1, -q^3, ..., -q^{2n+1}, ..., q^{2n+1}, ..., q^3, q^1, q^{-1}, q^{-3}, ..., q^{-2n+1}, ... ,
\]

all belonging to \(\mathbb{R}\). Consequently, we can state that instead of the individual \( f \) functions, we are integrating their equivalence classes consisting of the different \( f, g \), etc., which nevertheless coincide in the mentioned just now points.

\section{Integral equations}

The existence and uniqueness theorems regarding the solutions of differential equations play a key role in Physics since they allow the laws of nature to be formalized. In the works \cite{lavagno-gervino-09} it is shown that even if the special functions $E_q(x)$, $S_q(x)$, and $G_q(x)$ are solutions of the known Jackson \textit{ordinary differential} equation

\begin{equation*}
    \left[ D_x^{(q)} \right]^2 f(x) + a f(x) = 0,
\end{equation*}

the proof has been done by inspection.
But it seems there exist no said theorems and there are no algorithms for its solution in general. Although there are formulae, that link the Jackson derivatives with their respective Jackson integrals, namely

\begin{equation*}
    D_x^{(q)} \int_0^b f(t)_q dt = f(x),
\end{equation*}

and

\begin{equation*}
    \int_0^b D_x^{(q)}[f(x)]_q dx = f(b),
\end{equation*}

(here again we differ with \cite{lavagno-gervino-09} since, $f(0)$ does not exist in the above formula).
Just as for the antiderivative one can write

\begin{equation*}
    \int f(x) dx = (q^{-1} - q) \sum_{n=0}^{\infty} q^{2n+1} x f(q^{2n+1}x) + \text{const},
\end{equation*}

the problem is more attackable at level integral equations.
Let us assume that for the equation concerning $h(x)$ ($F(b)$ is a known function)

\begin{equation}
    \int_0^b h(x)_q dx = F(b),
    \label{eq:11}
\end{equation}

there are two solutions, namely $f(x)$ and $g(x)$. Replacing them and subtracting we obtain

\begin{equation*}
    b (q^{-1} - q) \sum_{n=0}^{\infty} q^{2n+1} [f(q^{2n+1}b) - g(q^{2n+1}b)] = 0,
\end{equation*}

for any allowed $q$. Since the sequence $q^3, q^5, \dots$ consists of independent $q$ functions we have

\begin{equation*}
    f(q^{2n+1}b) = g(q^{2n+1}b),
\end{equation*}

for $n = 0,1,2,\ldots$ or, in other words, the equivalence class of functions $\{ f(x) \}$ defined by us is the unique solution of Eq.~(\ref{eq:11}).

The proof of existence is reduced to the possibility of finding a class of functions among the elements of $S$ whose representatives would satisfy Eq.~(\ref{eq:11}). Differentiating Eq.~(\ref{eq:11}), taking in account we will find ourselves facing a problem of the Green function theory type \cite{Vladimirov-1971}.

\section{Discussion and conclusions}

The deformed algebras applicability in Quantum Physics inevitably involves the Jackson calculus. As a finite increment calculation this has its strengths and weaknesses. We believe that first of all there is the advantage of being able to consider the space as a lattice, that is, by \textit{\textbf{quantas}}. If the series defining the Jackson integrals can be presented as Fourier expansions, it would be possible, by means of the inverse transformation of this, to work in a bounded impulse space. Such a procedure would eliminate many singularities, including the problems that Field Theory encounters at point 0. Another advantage is the possibility of making adjustments with the deformation parameter $q$.  However, in our view, there are certain difficulties in the way of applying deformed theories to Quantum Physics. It is extremely important that only commuting operators can represent the simultaneous measurable magnitudes. This allows to construct the appropriate bases to find the search averages. Unfortunately, the vanish deformed commutators do not ensure simultaneous measurements. And this raises the question of whether it is feasible to “quantize” with this type of commutators. 
Another obstacle which is necessary to overcome is, as we already mentioned, the adjusting of the interpretation of the operator $\hat{N}$ (see \eqref{psi-0}) since its treatment in \cite{lavagno-gervino-09} is not correct. Only one fractional number of quantas contradicts quantum theory itself. In conclusion, we mention that we have managed to make Jackson calculus compatible with the equipped Hilbert spaces, correcting some errors such as the aforementioned problem with the proof of series convergence for negative subscripts and with respect to the analogy of the fundamental theorem of calculus. All this makes it possible in the future to try to rigorously apply what has been developed to lattice field theories.

\end{document}